\documentclass[conference,letterpaper]{IEEEtran}
\usepackage{hyperref}
\usepackage{xspace}
\usepackage{amsmath}
\usepackage{amsfonts}
\usepackage{amsthm}
\usepackage{paralist}
\usepackage{color}
\usepackage{xcolor}
\usepackage{mathtools}
\usepackage{array}
\usepackage{tikz}
\usetikzlibrary{arrows}
\usetikzlibrary{patterns}

\usepackage{stackrel}
\usepackage{pstricks}
\usepackage{breakurl}
\usepackage{subcaption}
\usepackage{amssymb}

\usepackage[utf8]{inputenc}
\usepackage{pgfplots}

\usepackage{booktabs}
\usepackage{expl3}
\usepackage{collcell}

\usepackage{soul}
\usepackage{graphicx}
\usepackage{listings}
\usepackage{subcaption}

\newcounter{note}[section]

\newcommand{\secref}[1]{\mbox{Sec.~\ref{#1}}\xspace}

\newcommand{\lineref}[1]{\mbox{line~\ref{#1}}\xspace}
\newcommand{\linesref}[2]{\mbox{lines~\ref{#1}--\ref{#2}}\xspace}

\newcommand{\secrefstatic}[1]{\mbox{Sec.~{#1}}}
\newcommand{\figref}[1]{\mbox{Fig.~\ref{#1}}\xspace}

\newcommand{\tblref}[1]{\mbox{Table~\ref{#1}}\xspace}

\newcommand{\chaprefstatic}[1]{\mbox{Ch.~{#1}}\xspace}

\newcommand{\kilobytes}{\ensuremath{\mathrm{KB}}\xspace}
\newcommand{\megabytes}{\ensuremath{\mathrm{MB}}\xspace}
\newcommand{\gigabytes}{\ensuremath{\mathrm{GB}}\xspace}

\newcommand{\gigahertz}{\ensuremath{\mathrm{GHz}}\xspace}
\newcommand{\secs}{\ensuremath{\mathrm{s}}\xspace}
\newcommand{\millisecs}{\ensuremath{\mathrm{ms}}\xspace}

\newcommand{\angr}{angr\xspace}
\newcommand{\tsgx}{T-SGX\xspace}
\newcommand{\klee}{KLEE\xspace}
\newcommand{\sTwoE}{S$^2$E\xspace}
\newcommand{\dart}{DART\xspace}
\newcommand{\cute}{CUTE\xspace}
\newcommand{\mayhem}{Mayhem\xspace}
\newcommand{\llvm}{LLVM\xspace}
\newcommand{\qemu}{QEMU\xspace}
\newcommand{\exe}{EXE\xspace}
\newcommand{\clang}{Clang\xspace}
\newcommand{\unix}{Unix\xspace}
\newcommand{\linux}{Linux\xspace}
\newcommand{\cliver}{CliVer\xspace}

\newcommand{\syscall}[1]{\texttt{#1}}
\newcommand{\select}{\syscall{select}\xspace}
\newcommand{\fork}{\syscall{fork}\xspace}
\newcommand{\stdout}{\texttt{stddout}\xspace}

\newcommand{\instr}[1]{\texttt{#1}}
\newcommand{\xbegin}{\instr{xbegin}\xspace}
\newcommand{\xend}{\instr{xend}\xspace}
\newcommand{\xabort}{\instr{xabort}\xspace}
\newcommand{\jmp}{\instr{jmp}\xspace}
\newcommand{\pop}{\instr{pop}\xspace}
\newcommand{\ret}{\instr{ret}\xspace}
\newcommand{\call}{\instr{call}\xspace}
\newcommand{\regRax}{\instr{\%rax}\xspace}
\newcommand{\regRnine}{\texttt{\%r9}\xspace}
\newcommand{\popRnine}{\:\pop \texttt{\%r9}\:\xspace}
\newcommand{\greg}{\texttt{greg\_t * gregs}\xspace }
\newcommand{\temptwo}{\texttt{\%2}\xspace}
\newcommand{\tempfour}{\texttt{\%4}\xspace}

\newcommand{\sysName}{TASE\xspace}

\newcommand{\stride}{\ensuremath{s}\xspace}
\newcommand{\maxStride}{\ensuremath{\stride_{\mathsf{max}}}\xspace}
\newcommand{\minStride}{\ensuremath{\stride_{\mathsf{min}}}\xspace}
\newcommand{\completedBBs}{\ensuremath{c}\xspace}
\newcommand{\byte}{\ensuremath{b}\xspace}
\newcommand{\byteAlt}{\ensuremath{b'}\xspace}
\newcommand{\byteBuf}{\ensuremath{B}\xspace}

\newcommand{\tran}{\ensuremath{\mathsf{tx}}\xspace}
\newcommand{\tranAlt}{\ensuremath{\mathsf{tx}'}\xspace}
\newcommand{\inputParam}{\ensuremath{V}\xspace}
\newcommand{\inputParamAlt}{\ensuremath{V'}\xspace}
\newcommand{\thread}{\ensuremath{\mathsf{thd}}\xspace}
\newcommand{\threadAlt}{\ensuremath{\mathsf{thd}'}\xspace}

\newcommand{\msgIdx}{\ensuremath{i}\xspace}
\newcommand{\timestamp}{\ensuremath{t}\xspace}
\newcommand{\cost}[1]{\ensuremath{\mathit{cost}({#1})}\xspace}
\newcommand{\lag}[1]{\ensuremath{\mathit{lag}({#1})}\xspace}

\newcommand{\arrival}[1]{\ensuremath{\mathit{arr}({#1})}\xspace}

\newcommand{\gmail}{Gmail\xspace}
\newcommand{\openssl}{OpenSSL\xspace}

\newcommand{\sclient}{\texttt{s\_client}\xspace}

\newcommand{\heartbleed}{Heartbleed\xspace}
\newcommand{\tcpdump}{\texttt{tcpdump}\xspace}

\newcommand{\processorModel}{3.5\gigahertz Intel Xeon CPU E3-1240 v5\xspace}
\newcommand{\RAMAmount}{64\gigabytes}

\newcommand{\cliverConcPercent}{2.7\%\xspace}

\newcommand{\basicTaseGmailAvgCost}{0.009\secs}
\newcommand{\basicTaseGmailMedianCost}{0.007\secs}
\newcommand{\basicTaseGmailMaxCost}{0.142\secs}
\newcommand{\basicTaseGmailAvgLag}{1.739\secs}
\newcommand{\basicTaseGmailMedianLag}{0.606\secs}
\newcommand{\basicTaseGmailMaxLag}{8.219\secs}

\newcommand{\basicCliverGmailAvgCost}{0.033\secs}
\newcommand{\basicCliverGmailMedianCost}{0.006\secs}
\newcommand{\basicCliverGmailMaxCost}{2.116\secs}
\newcommand{\basicCliverGmailAvgLag}{4.654\secs}
\newcommand{\basicCliverGmailMedianLag}{4.720\secs}
\newcommand{\basicCliverGmailMaxLag}{17.880\secs}

\newcommand{\optTaseGmailAvgCost}{0.027\secs}
\newcommand{\optTaseGmailMedianCost}{0.022\secs}
\newcommand{\optTaseGmailMaxCost}{0.107\secs}
\newcommand{\optTaseGmailAvgLag}{0.219\secs}
\newcommand{\optTaseGmailMedianLag}{0.066\secs}
\newcommand{\optTaseGmailMaxLag}{1.787\secs}

\newcommand{\optCliverGmailAvgCost}{0.070\secs}
\newcommand{\optCliverGmailMedianCost}{0.018\secs}
\newcommand{\optCliverGmailMaxCost}{1.274\secs}
\newcommand{\optCliverGmailAvgLag}{1.138\secs}
\newcommand{\optCliverGmailMedianLag}{1.195\secs}
\newcommand{\optCliverGmailMaxLag}{4.210\secs}

\newcommand{\nativeBNTime}{0.064\secs}
\newcommand{\nativeshaTime}{0.210\secs}
\newcommand{\nativemdTime}{0.125\secs}

\newcommand{\ourBNTime}{0.756\secs}
\newcommand{\ourshaTime}{1.825\secs}
\newcommand{\ourmdTime}{1.619\secs}

\newcommand{\stwoeBNTime}{2.907\secs}
\newcommand{\stwoeshaTime}{3.850\secs}
\newcommand{\stwoemdTime}{9.599\secs}

\newcommand{\kleeBNTime}{182.747}

\newcommand{\kleeshaTime}{528.718\secs}
\newcommand{\kleemdTime}{1276.700\secs}
\newcommand{\maxStrideVal}{16\xspace}
\newcommand{\transBBSizeVal}{50\xspace}

\begin{document}

\title{\sysName: Reducing Latency of Symbolic Execution with Transactional Memory}
\date{}

\author{
  \IEEEauthorblockN{Adam Humphries\IEEEauthorrefmark{1},
  Kartik Cating-Subramanian\IEEEauthorrefmark{2}, and
  Michael K.\ Reiter\IEEEauthorrefmark{1}}
  \IEEEauthorblockA{\IEEEauthorrefmark{1}University of North Carolina at Chapel Hill}
  \IEEEauthorblockA{\IEEEauthorrefmark{2}University of Colorado -- Boulder (work performed at UNC-Chapel Hill)}
}

\maketitle
\thispagestyle{plain}
\pagestyle{plain}


\begin{abstract}
  We present the design and implementation of a tool called \sysName
  that uses transactional memory to reduce the latency of
  symbolic-execution applications with small amounts of symbolic
  state.  Execution paths are executed natively while operating on
  concrete values, and only when execution encounters symbolic values
  (or modeled functions) is native execution suspended and
  interpretation begun.  Execution then returns to its native mode
  when symbolic values are no longer encountered.  The key innovations
  in the design of \sysName are a technique for amortizing the cost of
  checking whether values are symbolic over few instructions, and the
  use of hardware-supported transactional memory (TSX) to implement
  native execution that rolls back with no effect when use of a
  symbolic value is detected (perhaps belatedly).  We show that
  \sysName has the potential to dramatically improve some
  latency-sensitive applications of symbolic execution, such as
  methods to verify the behavior of a client in a client-server
  application.
\end{abstract}

\section{Introduction}
\label{sec:introduction}

Since its introduction~\cite{Boyer:1975:Select, King:1976:Symbolic},
symbolic execution has found myriad applications for security analysis
and defense (e.g.,~\cite{Kruegel:2005:Mimicry, Brumley:2006:Vulnsig,
  Cadar:2006:EXE, Yang:2006:Maldisks, Costa:2007:Bouncer,
  Wang:2009:Genomic, Milushev:2012:Noninterference,
  Shoshitaishvili:2015:Firmalice, Pasareanu:2016:Side-Channel,
  Zhou:2018:Noninterference}), software testing
(e.g.,~\cite{Visser:2004:PathFinder, Godefroid:2005:DART,
  Sen:2005:CUTE, Tillmann:2008:Pex, Anand:2008:Demand,
  Pasareanu:2008:NASA, Godefroid:2012:SAGE, Cadar:2013:Survey}), and
debugging (e.g.,~\cite{Zamfir:2010:Exesyn, Yuan:2010:SherLog}).
Whereas regular, ``concrete'' execution of a program maintains a
specific value for each variable, symbolic execution allows some
``symbolic'' variables to be undetermined but possibly constrained
(e.g., to be in some range).  Upon reaching a branch condition
involving a symbolic variable, each branch is executed under the
constraint on the symbolic variable implied by having taken that
branch.  Any execution path thus explored yields a set of constraints
on the symbolic variables implied by having taken that path.  In an
example use case, these constraints could be provided to an SMT
solver~\cite{Monniaux:2016:Survey} to compute a concrete assignment to
the symbolic inputs that would cause that path to be executed.

When applied to testing, the speed of symbolic execution is typically
a secondary concern.  However, several security applications place
symbolic execution on the critical path of defensive response in
time-critical circumstances.  For example, some works
(e.g.,~\cite{Brumley:2006:Vulnsig, Costa:2007:Bouncer}) leverage
symbolic execution to generate vulnerability signatures upon detecting
an exploit attempt, and so the speed of symbolic execution is a
limiting factor in the speed with which vulnerability signatures can
be created and deployed to other sites.  Other examples are
intrusion-detection systems in which a server-side \textit{verifier}
symbolically executes a client program to find an execution path that
is consistent with messages received from the client, without knowing
all inputs driving the client (e.g.,~\cite{Cochran:2013:Cliver,
  Chi:2017:Cliver}).  If each message could be verified before
delivering it to the server, then the server would be protected from
exploit traffic that a legitimate client would not send (e.g.,
Heartbleed packets to an OpenSSL server~\cite{Chi:2017:Cliver}).
However, such tools are not yet fast enough to perform this checking
on the critical path of delivering messages to the server, reducing
them to detecting exploits \textit{alongside} server processing.

Conventional wisdom holds that SMT solving and state explosion are the
primary latency bottlenecks in symbolic execution.  However, the speed
of straightline, concrete execution has been found to be the primary
culprit in some contexts (e.g.,~\cite{Cochran:2013:Cliver,
  Yun:2018:Qsym}).  Most symbolic execution tools incur a substantial
performance penalty to straightline execution because they interpret
the program under analysis, even when it is performing operations on
concrete data. For example, Yun et al.~\cite{Yun:2018:Qsym} report
straightline-execution overheads of 3000$\times$ and 321,000$\times$
native execution speed for \klee~\cite{Cadar:2008:Klee} and
\angr~\cite{Shoshitaishvili:2015:Firmalice}, due to interpretation.
The need to interpret the program in these tools arises from the need
to track symbolic variables, to accumulate constraints on those
variables along each execution path, and to explore multiple execution
paths.  Even attempts to optimize symbolic execution when processing
only instructions with concrete arguments must typically incur
overheads due to lightweight interpretation; e.g.,
\sTwoE~\cite{Chipounov:2011:S2E} encounters an overhead of roughly
$6\times$ vanilla \qemu~\cite{Bellard:2005:QEMU} execution speed on
purely concrete data due its use of memory sharing between \qemu and
\klee.

In this paper, we provide a solution that supports fast native
execution of instructions with concrete values---while still using
interpretation to do the ``symbolic parts'' of symbolic execution---on
modern x86 platforms.  Our design and associated tool, called
\sysName\footnote{\sysName stands for ``Transactional Acceleration for
  Symbolic Execution''.}, accomplish this through two key innovations.
Though \sysName instruments the executable to test whether variables
are concrete or symbolic (like \exe~\cite{Cadar:2006:EXE}), our first
innovation amortizes the costs of these tests by batching many into a
few instructions.  To maximize the benefits of this amortization,
\sysName defers these checks to ensure that only variables actually
used are checked; this deferment, together with the amortization,
means that instructions may be concretely executed on symbolic
variables.  Therefore, a critical second innovation in \sysName is a
way of rolling back such erroneous computations so they have no
effect.  \sysName uses hardware transactions as supported by Intel TSX
extensions for this purpose, though as we will see, accomplishing with
low overhead is nontrivial.

This paper outlines the design of \sysName, and evaluates its
potential to accelerate symbolic execution of applications with small
amounts of symbolic state.  We first show where \sysName improves over
modern alternatives such as \klee and \sTwoE through a microbenchmark
comparison.  This comparison shows that while \sysName can perform
poorly relative to these alternatives for applications with large
amounts of symbolic data, it can perform much better than them when
the amount of symbolic data is small.

Second, we show how \sysName qualitatively improves the deployment
options for a specific defensive technique, namely behavioral
verification of a client program~\cite{Cochran:2013:Cliver,
  Chi:2017:Cliver} as introduced above.  Though Chi et al.\ were able
to show the verification of \openssl client messages in TLS 1.2
sessions induced by a \gmail workload at a speed that coarsely keeps
pace with these sessions~\cite{Chi:2017:Cliver}, their verification
was not fast enough to perform on the critical path of message
delivery.  We show that replacing the symbolic execution component of
their tool with \sysName substantially improves the prospects for
performing verification as a condition of message delivery.  More
specifically, we show that \sysName's optimizations reduce the
average, median, and maximum \textit{lag} suffered by any
client-to-server message by over 80\%, 94\%, and 57\%, where lag is
defined as the delay between arrival of a message to the verifier and
its delivery to the server after verification completes.  In doing so,
\sysName brings these lags into ranges that are practical for
performing inline verification of TLS sessions driven by applications,
like \gmail, that are paced by human activity.

To summarize, our contributions are as follows:
\begin{compactitem}
\item We introduce a method to limit tests for determining whether
  variables are symbolic to only those variables that are actually
  used, and to batch many such tests into few instructions.  Since
  this deferred testing can result in our erroneously executing
  instructions on symbolic data, we show how transactional memory can
  be leveraged to undo the effects of these erroneous computations.
\item We detail the numerous optimizations necessary to realize the
  promise of this approach, in terms of achieving compelling
  performance improvements for some applications of symbolic execution.
  We show through microbenchmark tests where \sysName outperforms
  modern alternatives, \klee and \sTwoE.
\item We show that \sysName improves a specific defense using symbolic
  execution, namely behavioral verification~\cite{Cochran:2013:Cliver,
    Chi:2017:Cliver}, to an extent that qualitatively improves how
  such a defense can be deployed on TLS traffic.  Specifically, we
  show that \sysName reduces the costs of this defense to permit its
  application \textit{inline} for all but very latency-sensitive
  applications.  In doing so, \sysName enables preemptively protecting
  the server from exploits using this approach, versus its current
  ability to only detect malformed client messages alongside their
  processing by the server.
\end{compactitem}

The rest of this paper is structured as follows.  We discuss related
work in \secref{sec:related}.  We provide background and describe
challenges that we must overcome to realize \sysName in
\secref{sec:challenges}, and present the design of \sysName in
\secref{sec:design}.  We discuss additional aspects of \sysName's
implementation in \secref{sec:impl}.  \secref{sec:eval} contains an
evaluation of \sysName for a symbolic execution-based application,
and microbenchmarks.  We discuss limitations of \sysName in
\secref{sec:limitations} and conclude in \secref{sec:conclusion}.

\section{Related Work}
\label{sec:related}

In this section we outline prior work on symbolic execution systems
similar to ours, and work using Intel's Transactional Synchronization
Instructions.

\subsection{Symbolic Execution Engines}

Symbolic execution engines \dart~\cite{Godefroid:2005:DART} and
\cute~\cite{Sen:2005:CUTE} represent some of the earliest modern
attempts to mix concrete and symbolic execution
\cite{Cadar:2013:Survey}.  Their approach, called concolic testing,
analyzes a program by choosing an initial set of concrete input values
\inputParam to a given program.  The program is then executed with
instrumentation to determine when control flow instructions are
encountered, and constraints are accumulated at these branch locations
in terms of their relation to the concrete inputs.  After execution
with input \inputParam terminates or is suspended, the constraints
gathered from execution on \inputParam are analyzed to determine a new
set of concrete inputs \inputParamAlt to guide execution down a
different path.  The process can be run repeatedly until all paths are
explored, or until the tester wishes to cease path exploration.
Although we prioritize native execution in \sysName and mix concrete
and symbolic execution, our approach differs from concolic execution
in that we do not require entirely concrete inputs to drive symbolic
execution.  We also do not require re-execution of a program from a
new set of concrete values to reach different execution paths, as we
employ native forking (see \secref{sec:design:stateManagement}) to
explore different branches of program execution when control flow
depends on the value of a symbolic variable.

Rather than using a program's native execution state as its primary
representation, the \klee symbolic execution
engine~\cite{Cadar:2008:Klee} instead analyzes a program by
interpreting its source code translated to \llvm IR.  \klee is deeply
optimized to minimize the cost of constraint solving by caching
previous query results, applying normalization to constraints and
queries to facilitate comparisons between expressions, and analyzing
queries to determine subexpressions which may have already been
solved.  \klee is also structured to explore multiple program paths
within a single process.  By doing so, \klee is able to closely guide
state exploration with heuristics chosen to prioritize code coverage
or search for specific bugs or problematic behavior.  \klee also
implements software based copy-on-write to more efficiently manage the
symbolic states associated with different program paths.

\exe~\cite{Cadar:2006:EXE}, \mayhem~\cite{Cha:2012:Mayhem}, and
\sTwoE~\cite{Chipounov:2011:S2E} are symbolic execution engines that
use a program's native state as its principal representation.  \exe
analyzes a program by executing it natively and checking each use of a
variable against a map that indicates if the variable is symbolic.
Similarly, \mayhem uses dynamic taint
analysis~\cite{Newsome:2005:Taint} to detect instruction blocks that
touch symbolic data, while otherwise executing the program natively.
\exe and \mayhem also use forking to explore multiple execution paths,
i.e., forking the symbolic-execution process upon reaching a symbolic
branch, to allow the parent and child to explore the two possibilities
separately.  \sTwoE uses \qemu and \klee together to mix concrete and
native execution, and is the system most similar to \sysName.  \sTwoE
uses the virtualization and emulation tools within \qemu to perform
symbolic execution across user space and kernel space
boundaries~\cite{Chipounov:2011:S2E}.  \sTwoE also uses an emulated
MMU that checks each byte during access in concrete execution mode to
determine if control must transfer to the \klee-based
interpreter~\cite{Chipounov:2012:S2E}.  While we build on their
techniques for sharing symbolic and concrete state, \sysName is built
to prioritize and optimize native execution using new transactional
machine instructions and symbolic-state detection mechanisms detailed
in \secref{sec:design}.  For detecting symbolic state, \sysName does
not solely rely on the bitmap lookup techniques used in \exe and
\sTwoE, and \sysName incurs no virtualization or dynamic binary
translation overheads when executing code natively.

\subsection{Intel TSX}

Intel's Transactional Synchronization Instructions (TSX) were
originally introduced to speed up concurrency in multithreaded
applications~\cite[\chaprefstatic{16}]{Intel:2019:Manual}.  However,
TSX instructions have been repurposed for security defenses
(e.g.,~\cite{Shih:2017:T-SGX, Chen:2017:DejaVu}) and attacks
\cite{Jang:2016:DRK, Disselkoen:2017:PRIME}, as well.  Similarly,
\sysName uses the transactions produced by TSX in an unorthodox way.
Specifically, \sysName uses transactions to speculatively execute
regions of code natively during symbolic execution, aborting the
transaction if symbolic data is encountered.  Key challenges for
implementing this strategy are presented in
\secref{sec:challenges:tsx}.

\section{Background and Challenges}
\label{sec:challenges}

Our work to optimize symbolic execution for latency-sensitive
applications required us to build on research from seemingly unrelated
topics.  In this section we briefly cover necessary background and key
challenges that we address in \sysName, pertaining to executing
concrete operations natively but safely during symbolic execution
(\secref{sec:challenges:concrete}) and leveraging Intel TSX in this
context (\secref{sec:challenges:tsx}).

\subsection{Concrete Operations in Symbolic Execution}
\label{sec:challenges:concrete}

Past works (e.g.,~\cite{Godefroid:2005:DART, Cadar:2006:EXE,
  Chipounov:2011:S2E, Cha:2012:Mayhem}) have recognized the
significance of enabling native execution for entirely concrete
computations in symbolic execution engines.  However, the overwhelming
amount of such concrete operations present in some of our target
applications necessitate more aggressive optimizations in \sysName.
For example, in Chi et al.'s verification of \openssl
traffic~\cite{Chi:2017:Cliver}, which we explore as an application of
\sysName in \secref{sec:eval:cliver}, fewer than \cliverConcPercent of
instructions executed operate on symbolic data, even after extensive
protocol-specific optimizations to eliminate unnecessary concrete
operations (described as the \textit{optimized} configuration in
\secref{sec:eval:cliver:setup}).  To enable inline operation of this
verifier, it is thus necessary that concrete operation be optimized as
much as possible.

To do so, \sysName speculatively executes regions of code natively
within transactions, optimistically assuming that no operation in the
transaction reads or overwrites symbolic values.  Transactions are atomic,
and if any operation in a transaction reads or overwrites a symbolic
value, \sysName must abort the transaction and resume execution within
an interpreter---in our case, a modified version of the \klee
interpreter.  After the transaction completes within the interpreter,
\sysName resumes native execution if possible.

Separating concrete and symbolic execution into different execution
modes provided challenges for safely handling the symbolic expressions
the interpreter produces.  In particular, \sysName tracks symbolic
values by tainting them, specifically by augmenting \klee's
concrete/symbolic bitmaps with poison tainting and tracking.  This
required the design and verification of invariants to guarantee that
the transition between concrete and symbolic execution does not
unexpectedly overtaint or undertaint the program's execution with
symbolic values, invaliding the resulting analysis.  Moreover, because
execution no longer occurs entirely within an interpreter, there is a
risk that native execution might overwrite previously symbolic
variables with concrete data with no indication to the interpreter,
forcing us to adjust \klee's data structures to prevent such updates.

\subsection{Implementing Transactions with Intel TSX}
\label{sec:challenges:tsx}

A key contribution of our work is the use of Intel Transactional
Synchronization Instructions (TSX) to increase the speed of symbolic
execution.  We focus specifically on the use of the TSX Restricted
Transactional Memory instructions \xbegin and \xend.

Intel's TSX instructions were originally released to provide a
hardware-assisted tool for managing concurrency in a process.  A
thread \thread may speculatively attempt to acquire a shared resource
by using an \xbegin prior to entering the critical section.  \xbegin starts
a transaction in which any modifications to memory or registers made
by \thread are either entirely committed at the end of the transaction
(signified by \xend) or entirely discarded, at which time control for
\thread may transfer to a fallback path with simpler locking
primitives (e.g., a spin lock).  In other words, the transaction is
atomic.

Should another thread \threadAlt attempt to enter the critical section
and modify the shared resource while \thread is also altering the
resource in the transaction, one or both of the transactions will
abort and roll back~\cite[\chaprefstatic{16}]{Intel:2019:Manual}.
Transactions are rolled back when conflicts over shared resources are
detected between the read and write sets of \thread and \threadAlt,
potentially allowing both threads to operate in the critical section
simultaneously if \thread and \threadAlt do not read or write the same
shared data.  Conflicts in the read/write sets of \thread and
\threadAlt are detected by the cache coherence protocol, and enabling
concurrency with TSX can potentially outperform other locking methods
which categorically prevent multiple threads from executing in the
critical section concurrently, even if no conflicting memory accesses
would have occurred~\cite[\chaprefstatic{16}]{Intel:2019:Manual}.

Intel's transactional execution instructions provide the basis for our
speculative execution scheme.  The application of the transactions to
create a fast path, while conceptually simple, requires a large number
of details to be addressed.  First, as noted by Shih et
al.~\cite{Shih:2017:T-SGX}, forcing a program to execute entirely
within transactions introduces substantial challenges.  Placing each
basic block from the program within a single transaction introduces an
overhead of roughly $8\times$ native execution, and transaction size
is limited by cache size and associativity. Further complicating
matters, transactions may abort due to asynchronous interrupts, are
never guaranteed to commit, and must be carefully started and
committed to avoid nesting.

Second, our speculative native execution scheme requires an efficient
mechanism to abort transactions that encounter symbolic data.
Ideally, individual bytes containing symbolic values could be marked
as inaccessible by the OS (e.g., via page permissions) or a low-level
hardware mechanism (e.g., via debug registers) so as to force any
transaction accessing the byte to roll back.  Unfortunately, the large
granularity of page-level permissions and the scarcity of debug
registers limit the effectiveness of these solutions.  Another option
is to inject instrumentation into the program to query a lookup table
on each byte access (cf.,~\cite{Cadar:2006:EXE}); however this
approach incurs a performance penalty for additional read operations
and compare operations, may clobber the FLAGS register depending on
its implementation, and also impacts the number of operations that may
be placed within a single transaction.  \secref{sec:design} contains
our approach for overcoming these challenges.

\begin{figure}
\includegraphics[width=1.225\columnwidth]{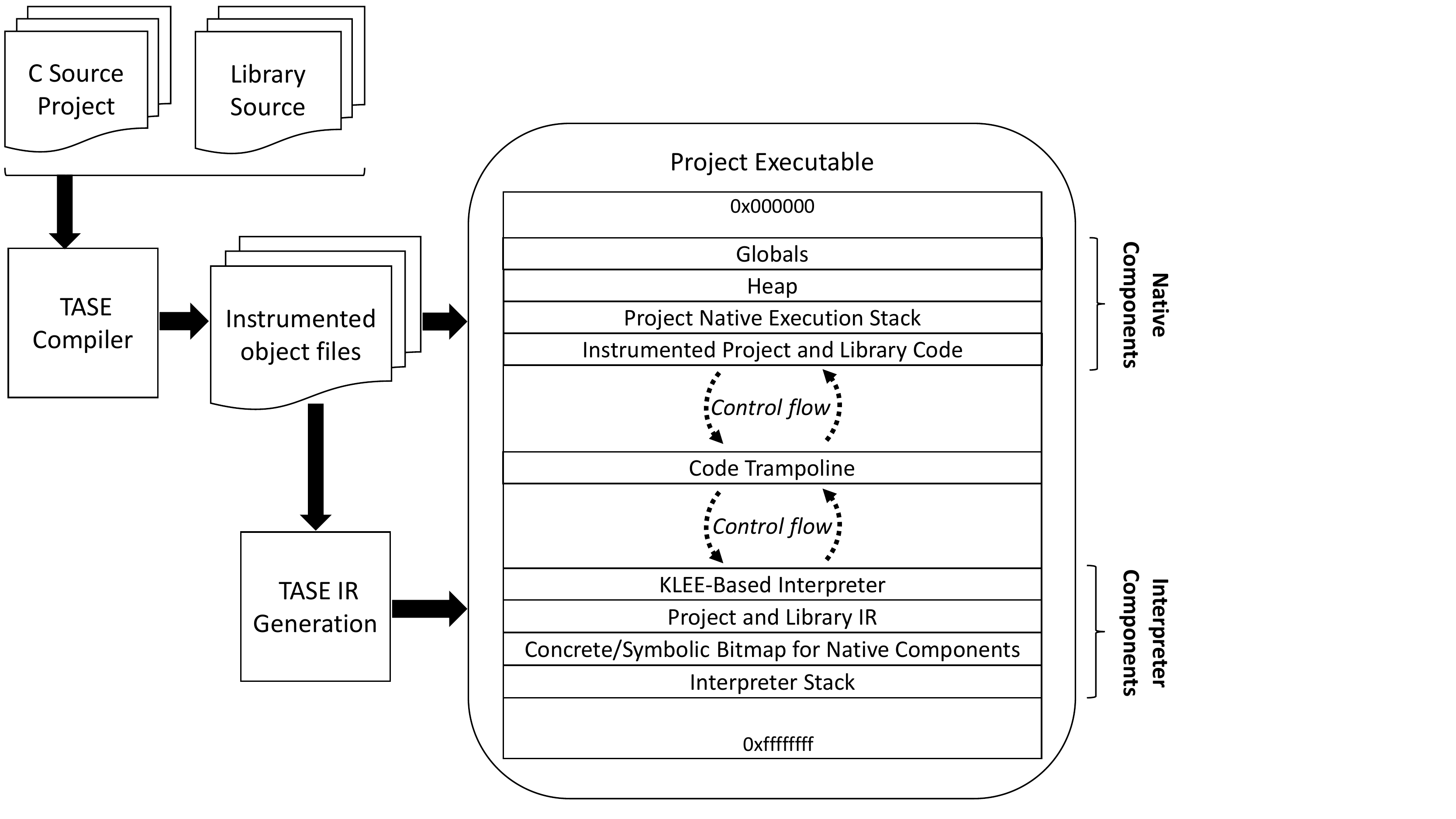}
\caption{High-level structure of \sysName.  \sysName comprises
  components labeled ``\sysName Compiler'' and ``\sysName IR
  Generation''.  \sysName generates a project executable containing
  native and interpretable representations of the project source, and
  that switches between these representations through a code
  trampoline to which control flows after native execution of a
  project basic block and after interpretation of a project basic
  block that leaves no symbolic values in the emulated registers.}
\label{fig:structure}
\end{figure}

\section{Design}
\label{sec:design}

In this section, we outline the design of \sysName.  We begin by
describing the overall architecture of \sysName, and follow with
descriptions of the system's transactional execution; its poison
checking scheme for detecting memory accesses of symbolic values; its
method of interpretation; and its mechanisms for managing state
exploration.

\subsection{Structure of \sysName }
\label{sec:design:structure}

In \sysName, we provide a symbolic execution system designed to
rapidly symbolically execute user-space programs with small amounts
of symbolic data.  At its core, \sysName provides a ``fast path''
and ``slow path'' for handling concrete and symbolic operations,
respectively, as it executes an application (henceforth referred
to as the \textit{project}).  \figref{fig:structure}
shows a simplified overview of these two primary components.

\sysName requires C source code to execute a project, including source
code for any C libraries the project will use.  The ``fast path'' for
native execution described earlier is an instrumented, binary x86
version of the project (and any libraries it uses) produced by
compiling the project's source code with our custom LLVM \sysName
compiler.  Crucially, \sysName executes within this instrumented
native execution path as the rule rather than the exception.  By
instrumenting loads and stores and inserting jumps to a code
trampoline (cf.,~\cite{Luk:2005:Pin, Shih:2017:T-SGX}) with
transactional instructions around basic blocks, \sysName enables
speculative native execution.  \sysName uses a \textit{poison} (or
\textit{sentinel}) value to mark bytes as containing symbolic values
while executing the project.  While executing code natively within a
transaction, values read and overwritten are recorded and checked en
masse with SIMD instructions at the end of a transaction.  If the
poison value was read or overwritten, the transaction is aborted and
all state changes performed during the transaction are undone; details
are provided in \secref{sec:design:transactionalexecution}.

If \sysName is unable to complete a transaction natively, control
transfers via a context switch from the trampoline to the ``slow
path'', our \klee-based interpreter.  The interpreter is responsible
for executing the target binary until another transactional entry
point is reached, at which time the target's execution might begin
again concretely.

Context switching between the interpreter and native execution in
\sysName closely resembles that in \sTwoE~\cite{Chipounov:2011:S2E}.
The interpreter and native execution share a common address space, and
a context switch from native execution to the interpreter occurs by
snapshotting the current state of the general purpose registers
(GPRs).  The interpreter then uses this snapshot to model each x86
instruction's effects on main memory and a simulated copy of the GPRs,
which is restored for concrete execution after a transactional
boundary is reached and symbolic values no longer reside within the
GPRs.

Symbolic data---including values used for altering control flow---are
exclusively handled by the interpreter, which may also fork state to
explore new execution paths.  Our forking mechanism uses the native
\unix \fork system call to explore execution paths, similar to the
techniques used in \exe~\cite{Cadar:2006:EXE}; we include more details
in \secref{sec:design:stateManagement}.  We discuss the mechanisms for
detecting usage of symbolic data during native execution in
\secref{sec:design:poison}.

\subsection{Transactional Execution}
\label{sec:design:transactionalexecution}

In order to mitigate the cost of interpreting instructions with
concrete operands, \sysName instead executes these instructions
natively within TSX transactions.  Our strategy is to speculatively
execute the target program natively for as many transactions as
possible, and abort a transaction if an ``unsafe'' operation occurs
that requires special handling of symbolic data via interpretation.
\sysName requires access to source code, and emits instrumented
machine code for the program along with the symbolic execution
components and interpreter in a single executable.

In a previous work that executed software in transactions for a
different purpose, \tsgx~\cite{Shih:2017:T-SGX} uses the \clang \llvm
back-end to conservatively estimate the read and write sets of
instructions and cache-way usage at compile time to efficiently group
together a large number of instructions in a single TSX transaction.
Their technique helps to maximize the number of instructions in a
transaction to amortize the overhead required for setting up and
committing or rolling back a transaction.
Unlike \tsgx, \sysName does not statically determine the number of
instructions to place in a transaction.  Our evaluation of \openssl
verification (see \secref{sec:eval:cliver}) revealed the need to
efficiently instrument code that frequently included variable-sized
loops and function pointers, both of which make effective compile-time
instrumentation challenging.  Using a custom \clang \llvm back-end,
\sysName injects trampoline jumps around basic blocks and dynamically
determines the boundaries for closing and opening a transaction at
runtime.

For any transaction, let its \textit{stride} denote the number of
basic blocks attempted within the transaction.  In our present
implementation, we currently use a transactional batching policy in
which the stride of each transaction, by default, is set to a constant
\maxStride; in our evaluation in \secref{sec:eval}, \maxStride is set
to \maxStrideVal.  If a transaction \tran aborts, then one possibility
would be to trap to the interpreter and simply interpret through the
whole aborted transaction.  However, a more refined approach that
leverages the \textit{reason} for the abort can optimize execution
considerably.

If \tran aborts due to reading or overwriting a poisoned memory location
(see \secref{sec:design:poison}), then \tran is aborted using an
\xabort instruction.  This instruction permits information about the
abort to be conveyed to the abort handler in a register.  We use this
facility to convey the number \completedBBs of basic blocks that were
successfully executed in the transaction (without detecting poison)
before the one where poison was encountered, which is tracked in a
counter updated by the trampoline.  In this case, \sysName attempts
another transaction \tranAlt beginning at the same place as \tran, but
with a stride of \completedBBs.  If \tranAlt completes successfully,
then the interpreter is invoked to interpret through the next basic
block (where poison is known to appear), and native execution is
resumed afterward, if possible.

If \tran aborts for another reason, then it is generally necessary to
interpret through the basic block where the abort occurred
(see~\cite[\secrefstatic{16.3.8.2}]{Intel:2019:Manual}).  For example,
if \tran aborted due to triggering a page fault, then it will likely
trigger the page fault again if retried in full~\cite{Kleen:2014:TSX}.
\sysName thus attempts to natively execute as many of the basic blocks
in \tran as possible while incurring few transaction aborts, before
leveraging the interpreter to interpret through the basic block that
induced \tran to abort.  \sysName does this using the following logic
(inspired by binary search), assuming the stride of \tran is
\maxStride:

\begin{compactenum}
\item $\stride \gets \maxStride/2$.
\item While $\stride \ge \minStride$ do:
  \begin{compactenum}
  \item Attempt a transaction \tranAlt of \stride basic blocks.
    \label{step:tranAlt}
  \item $\stride \gets \stride/2$ (regardless of whether \tranAlt
    aborted and, if so, the reason for the abort).
  \end{compactenum}
\item Trap to the interpreter and have it interpret through
  \minStride basic blocks. \label{step:interpret}
\end{compactenum}

After step~\ref{step:interpret}, \sysName resumes native execution
with its default stride of \maxStride.  Note that each \tranAlt of
stride \stride in step~\ref{step:tranAlt} will either advance the
program counter past the \stride blocks attempted if \tranAlt does
not abort, or will leave the program counter unchanged if \tranAlt
aborts.  The logic above attempts to ensure that if the condition
that induced \tran to abort is persistent (e.g., a page fault
incurred during a particular basic block), then the troublesome
basic block is interpreted in step~\ref{step:interpret}.

\subsection{Poison Checking}
\label{sec:design:poison}

In order to prevent native transactions from interacting
with symbolic information in an ``unsafe'' way, we implement a poison
checking scheme.  One such ``unsafe'' interaction we prohibit is the
loading of a symbolic variable into a register; this allows us to
assume that arithmetic performed within the general purpose registers
cannot access symbolic values and therefore needs no additional
instrumentation.

Broadly speaking, on a byte-level basis each memory location that
contains a symbolic value or expression is ``poisoned'' with a
reserved numeric value.  Reads and writes within a native transaction
are instrumented at compile time to store the values read and
overwritten in reserved SIMD and general purpose registers. At the end
of each basic block and prior to a transaction's commitment, the
values in the registers are tested in bulk to determine if a poison
value was read or overwritten.  If a poison value is found in the SIMD
registers, the transaction is aborted.

In order to make this scheme sound and efficient, several
implementation refinements were required.  First, we implement
poisoning on an aligned two-byte basis.  Byte-level poisoning would
potentially result in a large number of ``false positives'' in which a
native transaction reads a value concretely that, by coincidence,
matches the poison value.  Consequently, if any single byte \byte
needs to be marked as symbolic by the interpreter, we poison the
2-byte-aligned buffer \byteBuf containing \byte.  If \byteBuf also
contains a concrete byte \byteAlt, then we mark \byteAlt as symbolic
as well, but with a constraint that \byteAlt equals its concrete value
prior to \byteBuf's poisoning.  Of course, the risk of false positives
could be reduced further by poisoning 4-byte buffers, though we have
found 2-byte poisoning to suffice so far.

Second, to prevent natively executing instructions that read from or
write to memory locations containing symbolic data, we copy reads and
memory values prior to writes to our reserved SIMD and general purpose
registers during native execution.  These read and overwritten values are later
checked at the end of the basic block to determine if an instruction
could have operated on symbolic data.  If so, the transaction rolls
back and the interpreter handles the transaction, referencing its
internal bitmap indicator to determine if memory operands are concrete
or symbolic.  Compared to instruction-by-instruction instrumentation,
batching poison checks at the the end of the basic block in SIMD
registers reduces the total number of instructions required to perform
the poison checks, and simplifies the process of verifying that
instrumentation checks do not clobber the FLAGS register.

Third, to ensure that control flow within a transaction containing one
or more instructions operating on poisoned data reaches the SIMD
checks and aborts, we add additional instrumentation before indirect
control flow instructions which allow jumps to arbitrary destinations.
Without such a safeguard, our belated poison checking scheme could
allow poison-dependent indirect control flow arithmetic (e.g., jump
table calculations and function pointers) to erroneously transfer
control to destination addresses computed with the poison sentinel
value, thereby circumventing the poison checking logic.  Specifically,
we add an additional trampoline jump to the poison checking logic
before instructions that jump to an operand address (e.g., \call
\regRax, \jmp \regRax), effectively placing the instructions in a
separate basic block and preventing their execution if the operand is
symbolic; as stated earlier, control flow between basic blocks traps
to our interpreter if symbolic taint enters a register.  Similarly, if
the indirect control flow instruction performs a jump to an address stored
in memory pointed to by its operands (e.g., \ret), we inject an
additional poison check for the destination address and a jump to the
batched poison-checking logic before the instruction is executed.
Fortunately, assuming access to the source code for target
applications in \sysName and restricting our custom compiler's
instruction selection simplifies the task of instrumenting indirect
control flow instructions.

The above design points help to ensure that native execution does not
interact with symbolic values in any way, including ``clobbering''
writes to symbolic variables that would, in effect, concretize them
without notifying the interpreter.  Our ``in-place'' poison-checking
scheme along with \klee's symbolic bitmap indicator provides a
mechanism for accomplishing this.

The design choices for our poisoning scheme were made to minimize the
cost of instrumentation.  Whenever possible, our instrumentation to
save values read from or overwritten in memory is inserted into the target
code to prevent any additional reads from or writes to memory.  With the
help of alignment guarantees from the compiler, many instructions
reading data larger than a byte can be instrumented by moving the
data read from a general purpose register to a reserved
register, where it is later checked en masse with other data values.
Taint trackers such as Minemu~\cite{Bosman:2011:Minemu} use similar
techniques to reserve SIMD registers for instrumentation purposes.

\subsection{Interpretation}
\label{sec:design:interpretation}

Should a native transaction encounter symbolic data, control flow
in \sysName transfers to a \klee-based interpreter.  Given the
representation of the project as a set of x86 registers and an
address space, the interpreter executes instructions until a
transactional entry point is reached (i.e., an instruction
corresponding to the code trampoline) and the registers contain
no symbolic data.  The interpreter tracks symbolic data on a
per-byte basis.

Invoking the interpreter requires saving a snapshot of the GPRs
as they appeared at the end of the last successful transaction;
crucially, \sysName does not require that main memory is snapshotted
or copied during the context switch.
In the interpreter, reads and writes to main memory are performed
directly on the addresses being read from or written to.  However,
changes to the simulated x86 registers in the interpreter must be
faithfully tracked so that native execution can be resumed by a
context switch after interpretation has completed.  As \klee
interprets \llvm IR, we provide \llvm IR representations of each x86
machine instruction within a given transaction to preserve the
semantics of the program and produce a system state that may be
restored for native execution; see~\figref{fig:popRnine:CModel}
and~\figref{fig:popRnine:IRModel} for an example of how the x86
instruction \popRnine is modeled.  In order to avoid directly writing
\llvm IR, our method for producing an \llvm IR model for each x86
instructions is to write the state changes performed by the
instruction in C (as in \figref{fig:popRnine:CModel}), and use
\clang (\url{https://clang.llvm.org}) to emit \llvm IR (as in
\figref{fig:popRnine:IRModel}).  Note that the interpretation of
\popRnine in~\figref{fig:popRnine:CModel} is modeled as the execution
of a function.  The function takes a context containing a set of
simulated general registers (\greg), copies the value in main memory
pointed to by the simulated stack pointer to the interpreter's
simulated register \regRnine (\lineref{line:cmodel:rnine} in
\figref{fig:popRnine:CModel}), and increments the simulated stack and
instruction pointers (\linesref{line:cmodel:rsp}{line:cmodel:rip} in
\figref{fig:popRnine:CModel}).

Equivalent \llvm IR is provided in~\figref{fig:popRnine:IRModel}.
Note that the offsets of the stack pointer, \regRnine, and the
instruction pointer in the \greg struct are 15, 1, and 16
respectively. The instructions on
\linesref{line:irmodel:getspbegin}{line:irmodel:getspend} retrieve the
address pointed to by our simulated stack pointer into temporary
variable \temptwo, load the value at that address into \tempfour,
and store the result into the interpreter's model of \regRnine in
\lineref{line:irmodel:storernine}.  In \lineref{line:irmodel:loadraw},
our interpreter directly reads from \sysName's virtual memory located
at the address specified our model of \regRnine in the \greg struct;
because of this, context switches between the interpreter and native
execution do not require expensive copy operations for memory other than
the saving and restoring of our simulated registers.  The instructions
on \linesref{line:irmodel:spincfirst}{line:irmodel:spincsecond}
increment our simulated stack pointer and correspond to
\lineref{line:cmodel:rsp} in~\figref{fig:popRnine:CModel}.  Similarly,
\linesref{line:irmodel:incipfirst}{line:irmodel:inciplast} increment
the simulated instruction pointer to point to the next opcode, as in
\lineref{line:cmodel:rip} of \figref{fig:popRnine:CModel}.

\lstdefinestyle{customc}{
  belowcaptionskip=1\baselineskip,
  breaklines=false,
  numbers=left,
  xleftmargin=-0.5em,
  language=C,
  showstringspaces=false,
  basicstyle=\scriptsize\ttfamily,
  numbersep=-1.75em,
  escapechar=^
}

\begin{figure}[t]
  \begin{subfigure}[b]{\columnwidth}
  \begin{lstlisting}[style=customc]
    void interp_pop_r9 (greg_t* gregs) {
      gregs[REG_R9] = *(greg_t*)gregs[REG_RSP]; ^\label{line:cmodel:rnine}^
      gregs[REG_RSP] = gregs[REG_RSP] + 8; ^\label{line:cmodel:rsp}^
      gregs[REG_RIP] = gregs[REG_RIP] + 2; ^\label{line:cmodel:rip}^
    }
  \end{lstlisting}
  \caption{C model}
  \label{fig:popRnine:CModel}
  \end{subfigure}
  \begin{subfigure}[b]{\columnwidth}  
    \begin{lstlisting}[style=customc]
    define void @interp_pop_r9(i64* nocapture %gregs) #1 {
      %1 = getelementptr inbounds i64* %gregs, i64 15 ^\label{line:irmodel:getspbegin}^
      %2 = load i64* %1
      %3 = inttoptr i64 %2 to i64*
      %4 = load i64* %3 ^\label{line:irmodel:loadraw}^
      %5 = getelementptr inbounds i64* %gregs, i64 1 ^\label{line:irmodel:getspend}^
      store i64 %4, i64* %5  ^\label{line:irmodel:storernine}^
      %6 = add nsw i64 %2, 8 ^\label{line:irmodel:spincfirst}^
      store i64 %6, i64* %1 ^\label{line:irmodel:spincsecond}^
      %7 = getelementptr inbounds i64* %gregs, i64 16 ^\label{line:irmodel:incipfirst}^
      %8 = load i64* %7
      %9 = add nsw i64 %8, 2
      store i64 %9, i64* %7 ^\label{line:irmodel:inciplast}^
      ret void
    }
    \end{lstlisting}
    \caption{\llvm IR model}
    \label{fig:popRnine:IRModel}
  \end{subfigure}
  \caption{Models for interpreting \popRnine}
  \label{fig:popRnine}
\end{figure}

While the example described above and pictured in
\figref{fig:popRnine} is for a single instruction, \sysName instead
generates interpretable models for entire basic blocks of the original
project.  We elaborate further in \secref{sec:impl:ir}.

\subsection{State Management}
\label{sec:design:stateManagement}

In addition to managing the transition between native execution and
interpretation, \sysName must also handle the exploration of a
potentially large number of execution paths.  Handling this ``state
explosion'' problem is a crucial aspect of symbolic execution, and has
been a primary concern of many papers~\cite{Cadar:2013:Survey,
  Cadar:2008:Klee, Cha:2012:Mayhem,Chipounov:2012:S2E}.

In \sysName, multiple execution paths are explored in parallel by
using a native forking mechanism.  Unlike other systems that explore
multiple execution states within a single address space, \sysName is
unable to handle multiple execution states concurrently within a
single address space.  Attempting to explore states concurrently with
multiple threads on one address space could cause unintended
transactional aborts when threads access a common memory address.

Whenever the target program encounters a control flow instruction
(e.g., a \jmp or branch) that depends on a symbolic variable,
execution must revert to the interpreter.  After the interpreter takes
control, execution states are created corresponding to the different
possible destinations of the control flow instruction, and the \fork
system call is invoked.  The resulting two processes extend the
current execution in cases that the branch condition is true or false,
respectively.  We address indirect control-flow transfers dependent on
a symbolic variable by producing an execution state for each possible
destination.  \exe~\cite{Cadar:2006:EXE}
uses a similar mechanism to handle state exploration, and in both
\sysName and EXE this approach provides the benefit of hardware-based
copy-on-write to mitigate the cost of creating new processes.  Both
EXE and \sysName also have at least some cases in which state
exploration and path prioritization require child processes to halt
and wait for a central state management process to authorize further
execution.  This potentially introduces bottlenecks when many child
processes are exploring a large state space; however the
centralization of state management in a single process helps to
prevent ``fork bombing'' issues in which the machine hosting \sysName
is overwhelmed with too many processes.

Forcing each process following a \fork to signal back to the central
management process allows a variety of search heuristics to be
implemented by the central coordinator.  We intend to explore the use
of simple heuristics, such as breadth-first and depth-first search, as
well as ones tailored to particular applications.  For example, in
prior research on client behavior verification, Cochran et
al.~\cite{Cochran:2013:Cliver} leveraged the next message inbound from
the client to prioritize the order in which paths were explored to
identify a path consistent with that message having been sent next by
the client.  This prioritization was based on data collected from the
client program during a training phase.  In this approach, when a path
search reaches a symbolic branch, the central coordinator determines
which of the currently paused processes---i.e., either the two
resulting from this \fork, or another one---is on a path that is
``closest'' to one that, in training, could typically be used to
``explain'' the latest message received from the client.  That process
would then be signaled to continue its search until reaching the next
symbolic branch.  Of course, this prioritization is only an example
strategy, and we intend to explore others, as well.

\section{Implementation}
\label{sec:impl}

In this section we briefly discuss implementation details of \sysName.

\subsection{IR Generation}
\label{sec:impl:ir}

Like other symbolic execution engines, \sysName requires an intermediate
representation of code to perform symbolic execution.  Specifically,
\sysName uses \llvm IR to model each x86 instruction that potentially
touches symbolic data, as discussed in \secref{sec:design:interpretation}.

Crucially, unlike some other symbolic execution tools, \sysName
requires access to source code, from which \sysName produces an
instrumented executable using a custom compiler.  Controlling the
compiler allows us to selectively limit the pool of instructions
available to the \llvm backend's code-generation algorithms.  This
drastically simplifies the laborious task of producing IR models for
x86 instructions, at the cost of requiring source code.

Additionally, we use information provided by the \llvm backend during
compilation to record FLAGS-register liveness information around basic
blocks, which we use to periodically kill the FLAGS register.  This
benefits our execution in \sysName because it reduces the overall
amount of symbolic data the interpreter must handle, and, in certain
situations, allows the interpreter to more quickly produce a fully
concrete copy of its simulated GPRs needed to return to native
execution.

Because execution within a basic block in \sysName must occur either
entirely in the interpreter or natively for the duration of the basic
block, we employ an additional optimization to speed up interpretation.
We ``batch'' the IR for all x86 instructions in a basic block together
and invoke the interpreter to interpret whole basic block at once,
rather than doing so per instruction within the basic block.  In
practice, we observe that this optimization reduces the total size of
the \llvm interpretation bitcode by a factor of roughly three.
Assuming access to source and control over the compiler also helps
here; by disabling the selection of instructions that modify certain
flags bits (e.g., the direction flag used by string-manipulation
instructions), the overall size of the IR is reduced and more
opportunities to omit redundant flags computations appear.  Moreover,
we found that reducing instruction selection based on flags usage
offered opportunities to completely kill flags in certain cases after
control flow instructions were used, reducing the likelihood of
expressions ``snowballing'' together due to flags computations being
continuously OR'd together.

\subsection{Forking and Path Exploration}
As noted in \secref{sec:design:stateManagement},
we employ a native \unix \fork call to
explore multiple execution states in \sysName when execution
encounters a symbolic branch.  Execution in \sysName begins with a
central ``manager'' process forking off a child process to begin path
exploration of the project's code.  The manager uses signal-based
job-control mechanisms, shared memory, and system-level semaphores to
steer and control execution through different branches as a
pre-defined maximum number of worker processes execute in parallel.
If a worker encounters a symbolic branch, it halts execution until the
manager process determines what course of action to take.  Although
path selection and code coverage is vital for general symbolic
execution, we leave these for future work and instead focus on
optimizing workloads with low levels of symbolic taint.

Native forking in \sysName benefits from hardware-based copy-on-write,
but still incurs overhead; among other things, the \linux kernel
copies the parent process' page table entries for the
child~\cite{fork:2017}.  To reduce this cost, our experiments in
\sysName use the \linux transparent huge pages feature to reduce the
size of page table mappings without explicitly modifying the
applications.  The daemon used by the kernel to coalesce small
(4\kilobytes) pages into huge (2\megabytes) pages periodically
runs at a predefined interval; we experimentally determined that
10\millisecs appeared roughly optimal for our behavioral verification
application in \secref{sec:eval:cliver}.

\subsection{Transaction Sizing}
As discussed in \secref{sec:design:transactionalexecution}, the stride
of a transaction in \sysName is set to a constant \maxStride by
default; after executing \maxStride basic blocks, the transaction will
be closed.  A value \maxStride that is too small will hurt performance
by closing transactions more frequently than necessary, whereas a
value that is too large can incur a substantial performance penalty
when a transaction aborts, since all the work it performed will be
thrown away.  To maximize performance, \maxStride would ideally be
tuned per project and per platform, since the size of the L1 data
cache limits the amount of data that a transaction can read or write
and since the frequency at which symbolic data is accessed may vary
depending on the application.  In the future, we plan to explore
dynamically adjusting \maxStride based on runtime conditions, as well.
For the purposes of our evaluation in \secref{sec:eval}, we simply set
$\maxStride = \maxStrideVal$.

Because the basic block is the smallest granularity at which
transaction size can be controlled in \sysName, it is also necessary
that basic blocks be limited to a maximum size.  In our
present implementation, basic blocks are limited to \transBBSizeVal
instructions. Here again, the limit of \transBBSizeVal was chosen
experimentally; we plan to explore methods in future work to
automatically tune this constant.

\section{Evaluation}
\label{sec:eval}

In this section we measure \sysName's performance.  We first detail
\sysName's performance in a series of microbenchmarks in
\secref{sec:eval:micro}, and then we consider an application of
symbolic execution to validating the messaging behavior of a software
client in \secref{sec:eval:cliver}.  All performance experiments
described in this section were conducted on a computer with a
\processorModel processor and \RAMAmount of RAM.

\subsection{Microbenchmarks}
\label{sec:eval:micro}

In this section we report the results of various microbenchmarks to
compare \sysName to alternatives when executing on mostly concrete
workloads---the contexts for which \sysName was designed.  Our first
microbenchmark evaluations compared \sysName to native execution and
execution by \sTwoE and \klee,\footnote{We used Dockerized \klee built
  from a container retrieved on Oct 23, 2019
  (\url{https://klee.github.io/docker/}) and \sTwoE retrieved on July
  11, 2019 (\url{https://github.com/s2e/s2e-env.git}).} for three
programs: the first adds two concrete 10\megabytes integers
byte-by-byte with a one-byte carry, and the second and third are
sha256\footnote{\url{https://github.com/coreutils/gnulib/blob/master/lib/sha256.c}}
and
md5sum\footnote{\url{https://github.com/kfl/mosml/blob/master/src/runtime/md5sum.c}},
each applied to a concrete 44\megabytes file.  These programs were
compiled using \clang 7.1.0 with O2 optimization for the native and
\sTwoE targets, and with \clang 6.0.1 with O2 optimization for \klee
(as 6.0.1 was the \clang version included with \klee).  In contrast,
\sysName supports only a limited version of O1 optimization at the
time of this writing.

The results of these executions are shown in \tblref{table:concStats}.
\sysName was up to $13\times$ slower than native execution on these
benchmark programs.  However, \sTwoE was up to $77\times$ slower, and
\klee incurred overheads up to $10^4\times$ native execution.

\begin{table}
  \begin{center}
  \small
  \begin{tabular} {@{} l  r @{\hspace{0.9em}}  r @{\hspace{0.9em}} r @{\hspace{0.9em}} r @{}}
    \toprule
   \multicolumn{1}{c}{Microbenchmark} &  \multicolumn{1}{c}{Native} & \multicolumn{1}{c}{\sysName} & \multicolumn{1}{c}{\sTwoE} & \multicolumn{1}{c}{\klee}   \\
   \midrule
  BigNum addition (10\megabytes) & \nativeBNTime & \ourBNTime & \stwoeBNTime & \kleeBNTime  \\
  sha256 (44\megabytes) & \nativeshaTime & \ourshaTime & \stwoeshaTime & \kleeshaTime \\
  md5sum (44\megabytes) & \nativemdTime & \ourmdTime & \stwoemdTime & \kleemdTime  \\
  \bottomrule
  \end{tabular}
  \end{center}
\caption{Overhead on concrete computations.  All times in seconds.
  BigNum addition was performed byte-by-byte on two 10\megabytes
  integers.  Hashes were computed on a 44\megabytes file.}
\label{table:concStats}
\end{table}

\begin{figure}
  \centering
  \includegraphics[width=0.7\columnwidth]{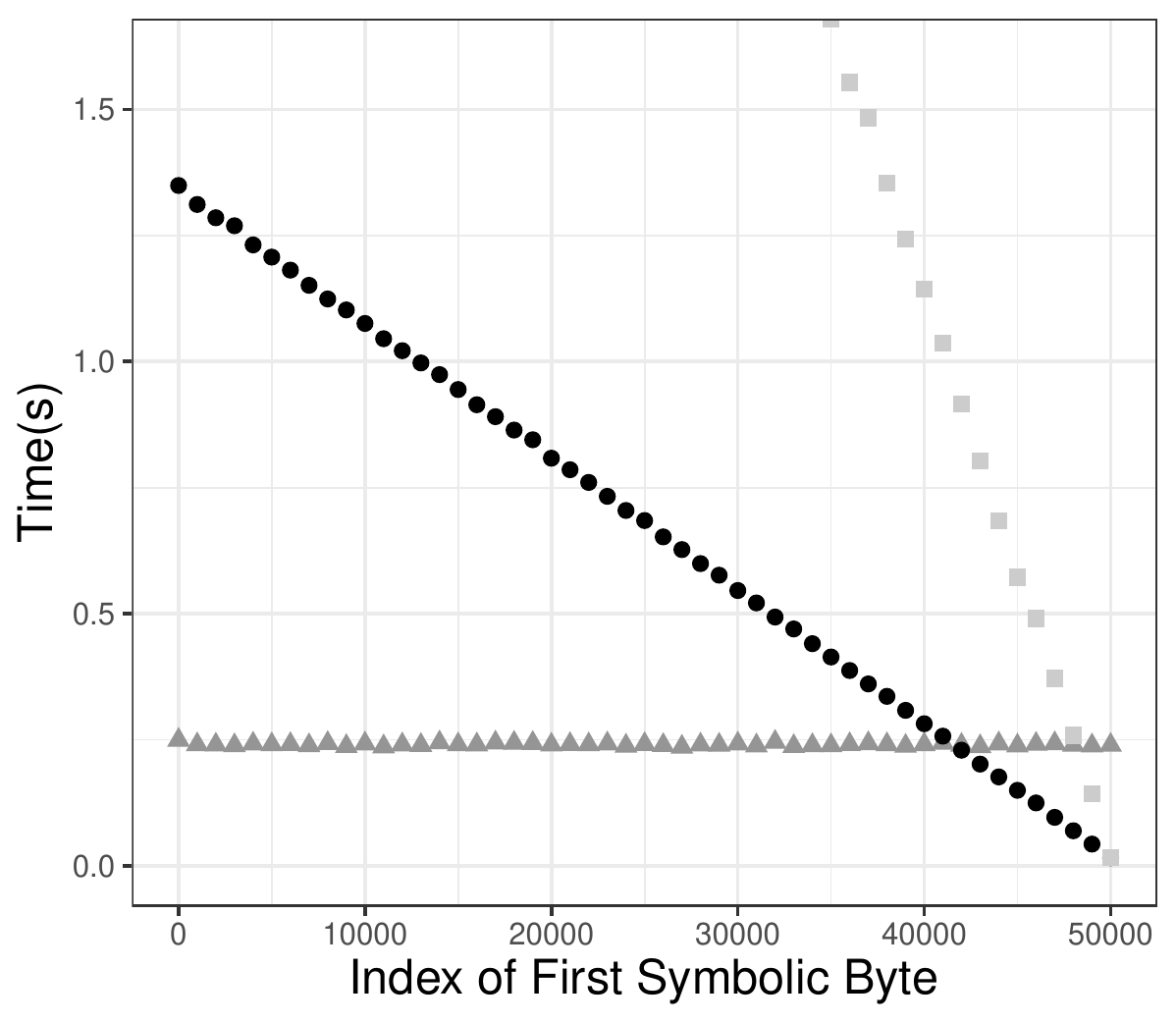}
  \caption{Average time to add two 50\kilobytes integers vs.\ index of
    first symbolic byte, for \sysName~($\bullet$),
    \klee~(\textcolor{gray}{$\blacktriangle$}), and
    \sTwoE~(\textcolor{lightgray}{$\blacksquare$}).  A higher index
    for the first symbolic byte indicates a higher fraction of
    concrete operations.  Each point is an average over five runs,
    with relative standard deviation $< 6\%$.}
  \label{fig:slidingScale}
\end{figure}

\sysName is tailored to executing projects with small amounts of
symbolic data, and so increasing the amount of symbolic data does
impact its performance.  \figref{fig:slidingScale} shows the
performance of byte-by-byte BigNum addition using the same code
represented in \tblref{table:concStats} (but only 50\kilobytes
operands), but with some byte marked symbolic.  Once this byte is
encountered, the carry byte becomes symbolic and remains so for the
rest of the computation; as such, the bytes of the sum tainted by the
symbolic carry byte are symbolic, as well.  This symbolic data did not
affect the performance of the BigNum addition in \klee
(\figref{fig:slidingScale}), since \klee interprets all project
instructions.  The performance of the BigNum addition in \sysName,
however, decayed as the first symbolic byte was encountered earlier in
the computation; only once $\approx 80\%$ of the BigNum addition was
performed concretely does \sysName outperform \klee.  The primary
reason for the loss of performance for larger amounts of symbolic data
is that \sysName interprets substantially more LLVM IR instructions to
model a restorable context for native execution (see
\secref{sec:limitations:instructions}).  In particular, \sysName
executed almost $8\times$ as many IR instructions as \klee in the
BigNum addition test when the index of the first symbolic byte was set
to zero.  \sTwoE exhibited similar trends as \sysName, eventually
becoming faster than \klee; however, its performance was much worse
than \sysName in this test, and became faster than \klee only once $>
95\%$ of the BigNum addition was concrete.

\subsection{Client Verification with \sysName}
\label{sec:eval:cliver}

The second evaluation for \sysName that we report is its use in a
framework for verifying client behavior in client-server
protocols~\cite{Chi:2017:Cliver}.  Numerous server exploits take the
form of a client sending messages to a server that no legitimate or
sanctioned client would send; e.g., Chi et al.\ observed that several
notable TLS exploits (e.g., \heartbleed, CVE-2014-0160) are of this
form.  To determine whether or not a client message could have
originated from an unmodified client TLS implementation, Chi et
al.\ detail a technique for symbolically executing \openssl's \sclient
and then solving to determine whether there exist inputs that could
have caused that implementation to produce the message sequence
received.  A message sequence for which no inputs can be found to
produce it indicates that the message sequence is inconsistent with
the claimed software client and so might represent an exploit.  The
Chi et al.\ framework is an extension of similar tools
(e.g.,~\cite{Bethea:2011:Server-side, Cochran:2013:Cliver}) adapted
specifically for cryptographic protocols like TLS: it leverages
knowledge of the TLS session key and symbolically executes the client
in multiple passes, skipping specified \textit{prohibitive functions}
(the AES block cipher and hash functions) until constraints generated
from observed client-to-server messages could fully concretize their
inputs.

This defense benefits from its generality (it needs no foreknowledge
of a vulnerability or exploit to detect an attack) and soundness, in
the sense that if it accepts a sequence of messages, then there are
inputs that could have caused the legitimate client to produce that
sequence.  However, to be used as an inline defense, it requires
symbolic execution to keep pace with the arrival of messages.  Below
we compare verification speeds using \sysName against the \klee-based
implementation originally produced by Chi et al., to which we refer
here as \cliver (for simply ``client verification'').  We show not
only that \sysName significantly improves performance, but that it
does so to an extent that permits this defense to reside on the
critical path of message processing for all but the most
latency-sensitive TLS applications.

The only changes we made to the \cliver tool for this evaluation was
to implement the following two optimizations for it, to make the
comparison to \sysName fairer since \sysName incorporates analogous
optimizations.  First, we changed how \cliver models the \select
system call, so that its return value indicates that \stdout is always
available (versus being symbolic).  \sclient writes the application
payload received from the server to \stdout, and so blocks if \stdout
is unavailable.  As such, this change has no effect on the message
sequence that could be received from \sclient; i.e., any message
sequence received in an execution where \stdout becomes unavailable is
a prefix of a sequence that could be received in an execution where it
remains available throughout.  This change does, however, relieve
\cliver from needing to explore the execution path in \sclient where
\stdout is unavailable, saving it the expense of doing so.

Second, when \cliver is seeking to verify message \msgIdx from the
client and reaches a send point when symbolically executing \sclient,
it must create and solve constraints reflecting message \msgIdx and
the path executed to reach that send point.  This produces an
unusually large number of relatively simple equality constraints
(i.e., one constraint per each byte of the message), many of which
contain a large number of XOR operations due to the choice of cipher
suite.  To more efficiently move the constraint information between
the interpreter and its solver, we alter the behavior of \klee's
independent constraint solver to send all constraints en masse rather
than one-by-one.  Moreover, though the SAT solver we use supports XOR
expressions~\cite{Soos:2009:Crypto}, we found it much more efficient
to rewrite these expressions to remove XORs before sending them to the
solver.  This optimization improves the performance of \cliver
considerably, and we leverage it in \sysName, as well.

\subsubsection{Experiment setup}
\label{sec:eval:cliver:setup}

Our evaluation used the same TLS 1.2 dataset used by Chi et
al.~\cite{Chi:2017:Cliver}.  This dataset includes benign traffic
captured by \tcpdump during a \gmail browsing session, and maliciously
crafted \heartbleed packets to simulate CVE-2014-0160.  The \gmail
data set was generated by sending and receiving emails with
attachments in Firefox over a span of approximately 3 minutes, and
included 21 independent, concurrent TLS sessions for a total of
3.8\megabytes of data.

We compared \sysName with \cliver in two configurations.  The first
presumes minimal protocol-specific knowledge or thus adaptation by the
party deploying the verifier.  In this \textit{basic} configuration,
each tool was provided a specification of the same prohibitive
functions, but otherwise the tool operated on the \gmail trace
unmodified.  Even in this configuration, however, we provided \cliver
with native implementations of these prohibitive functions, so that
even once their inputs had been concretized, they would be executed
natively (versus being interpreted), thus rendering our comparison
conservative.  The second, \textit{optimized} configuration
incorporated a range of protocol-specific optimizations.  In
particular, after the TLS 1.2 handshake, client-to-server and
server-to-client messages are independent of one another, and so
server-to-client messages were ignored when verifying the
client-to-server messages.  In addition, certificate verification was
elided, since the verifier, being deployed to protect the server,
trusts the server to send a valid certificate chain.

\subsubsection{Results}
\label{sec:eval:cliver:results}

When used as an inline defense against malicious traffic, the speed to
reach a true detection is arguably a secondary concern; the delay
imposed on attack traffic might be viewed more as a benefit than a
detriment.  Nevertheless, we used synthetic \heartbleed packets to
confirm that \sysName could determine these packets were not
consistent with the \openssl TLS client in only $178\millisecs$ from
the initiation of the connection (i.e., including the TLS handshake).

More critical is the delay that \sysName would impose on legitimate
traffic.  Here we report the \textit{cost} and \textit{lag} of
verification as defined by Chi et al.  For the \msgIdx-th message in a
TLS session, \cost{\msgIdx} is the processing time required to verify
message \msgIdx beginning from the symbolic state produced from
verifying through message $\msgIdx-1$.  \lag{\msgIdx} is the delay
between the receipt of message \msgIdx and when its verification
completes.  Note that $\lag{\msgIdx} \ge \cost{\msgIdx}$, and
$\lag{\msgIdx} > \cost{\msgIdx}$ if when message \msgIdx is received
by the verifier, the verification of message $\msgIdx-1$ is not yet
complete (and so verifying message \msgIdx cannot yet begin).

\tblref{tbl:gmailStats} gives coarse statistics for the cost and lag
of verifying all 21 TLS sessions in the \gmail trace, in the
\textit{basic} (left) and \textit{optimized} (right) configurations.
Interestingly, the median costs for \sysName and \cliver were very
similar, but the median lag for \cliver was up to $18\times$ larger (in
the \textit{optimized} configuration) than it was for \sysName.  The
cause was the messages that were most costly to verify, with costs in
\cliver roughly $12\times$ that in \sysName in the \textit{optimized}
configuration (and roughly $15\times$ that in \sysName in the
\textit{basic} configuration).  These greater costs caused the lag to
accumulate at various points in the trace, inducing an average \cliver
lag on the \textit{optimized} configuration of $> 1\secs$ and a
maximum lag of $> 4\secs$.  In contrast, the \sysName lag averaged
only $\approx 0.2\secs$ and incurred a maximum lag for any message of $<
2\secs$.  For a driving application like \gmail that is paced by human
activity, these lags may well be small enough to support the use of
\sysName as an inline defense.

\begin{table}
  \begin{center}
    \begin{tabular}{l@{\hspace*{3em}}rr@{\hspace*{3em}}rr}
    \toprule
  Configuration & \multicolumn{2}{l}{~~~~~~\textit{basic}} & \multicolumn{2}{l}{~~~\textit{optimized}} \\
  System        & \sysName & \cliver & \sysName & \cliver \\
  \midrule
  Average cost & \basicTaseGmailAvgCost & \basicCliverGmailAvgCost & \optTaseGmailAvgCost & \optCliverGmailAvgCost \\
  Median cost  & \basicTaseGmailMedianCost & \basicCliverGmailMedianCost & \optTaseGmailMedianCost & \optCliverGmailMedianCost \\
  Max cost     & \basicTaseGmailMaxCost & \basicCliverGmailMaxCost & \optTaseGmailMaxCost & \optCliverGmailMaxCost \\
  \midrule
  Average lag  & \basicTaseGmailAvgLag & \basicCliverGmailAvgLag & \optTaseGmailAvgLag & \optCliverGmailAvgLag \\
  Median lag   & \basicTaseGmailMedianLag & \basicCliverGmailMedianLag & \optTaseGmailMedianLag & \optCliverGmailMedianLag \\
  Max lag      & \basicTaseGmailMaxLag & \basicCliverGmailMaxLag & \optTaseGmailMaxLag & \optCliverGmailMaxLag  \\
  \bottomrule
  \end{tabular}
  \end{center}
\caption{Statistics for verification of benign \gmail traces}
\label{tbl:gmailStats}
\end{table}

\begin{figure}
  \centering
  \hspace{-0.75em}
  \begin{subfigure}[b]{.7\columnwidth} 
    \includegraphics[width=\columnwidth]{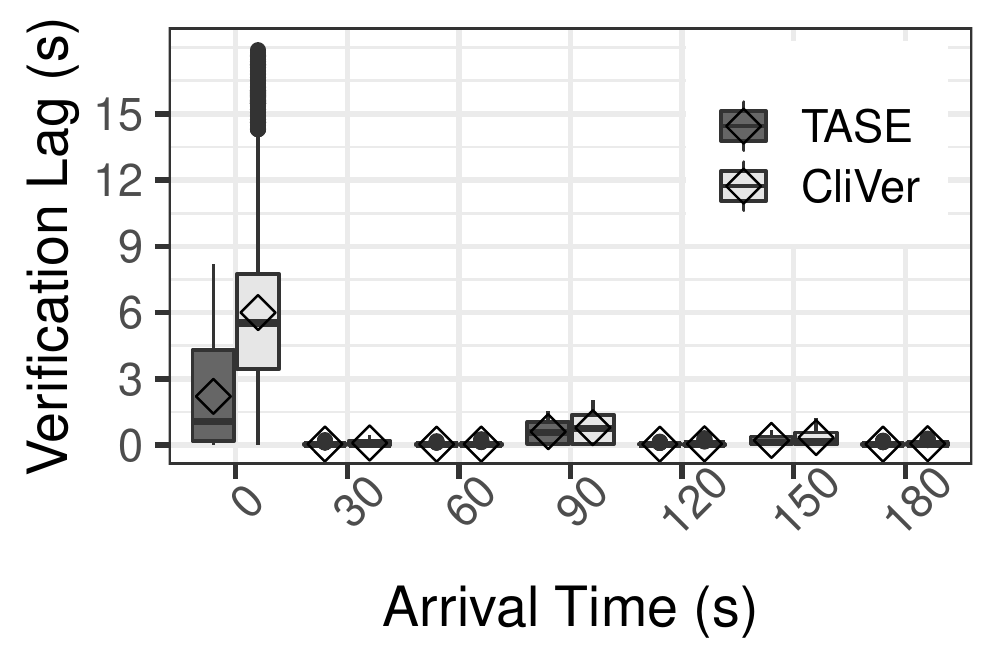}
    \caption{\textit{basic} configuration}
    \label{fig:lag:basic}
  \end{subfigure}
  \\[12pt]
  \begin{subfigure}[b]{.7\columnwidth}
    \includegraphics[width=\columnwidth]{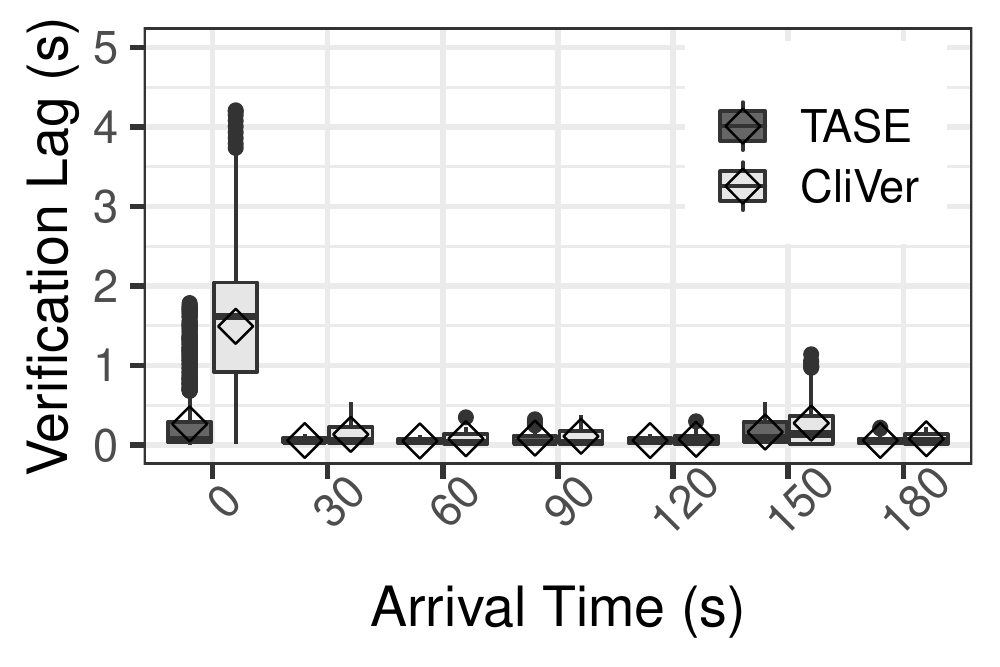}
    \caption{\textit{optimized} configuration}
     \label{fig:lag:opt}
  \end{subfigure}
\caption{Verification lag for 21 TLS sessions in the \gmail trace.
  The box plot at arrival time \timestamp includes $\{\lag{\msgIdx}:
  \timestamp \le \arrival{\msgIdx} < \timestamp + 30\secs\}$ across
  all 21 TLS sessions.  \figref{fig:lag:basic} shows lag in verifying
  the \gmail traces using \sysName and \cliver in a basic deployment
  without protocol-specific optimizations, and \figref{fig:lag:opt}
  shows lag in a configuration with optimizations leveraging protocol
  knowledge; see \secref{sec:eval:cliver:setup}.}
\label{fig:lag}
\end{figure}

A temporal view of the lag is pictured in \figref{fig:lag}, which
shows the distribution of lag across all 21 TLS sessions with
messages binned according to their arrival times, where
\arrival{\msgIdx} denotes the arrival time of message \msgIdx; for
example, the first bin contains the messages that arrived within the
first 30\secs of each of the 21 TLS sessions.  Arrival time is
measured relative to the start of the individual TLS session.  Within
each bin, a box-and-whisker plot shows the first, second, and third
quartiles, with the whiskers extended to $1.5\times$ the interquartile
range.  The average is shown as a diamond, and outliers appear as
individual points.

The lags for the \textit{basic} deployment lacking protocol-specific
optimizations are shown in \figref{fig:lag:basic}, and the lags for
the \textit{optimized} deployment leveraging protocol-specific
optimizations are shown in \figref{fig:lag:opt}.  Both \sysName and
\cliver suffered lag in the first 30\secs of each connection, though
\sysName' median lag in this interval was $\approx 20\%$ of \cliver's
in the \textit{basic} configuration, and even less in the
\textit{optimized} configuration.  Indeed, the 75\textsuperscript{th}
percentile of \cliver's lag in this first 30\secs exceeded essentially
all lags induced by \sysName in the same interval.  By the end of the
first 30\secs, both tools ``caught up'' and maintained lags capable of
sustaining interactive use until about 90\secs into the traces; at this
point, large server-to-client transfers caused \cliver to lag
considerably in the \textit{basic} configuration, while \sysName was
able to better keep up.  These lags were smaller in the
\textit{optimized} configuration, since server-to-client data messages
were ignored.

\begin{figure}
  \hspace{-0.75em}
  \begin{subfigure}[b]{.5\columnwidth} 
    \includegraphics[width=\columnwidth]{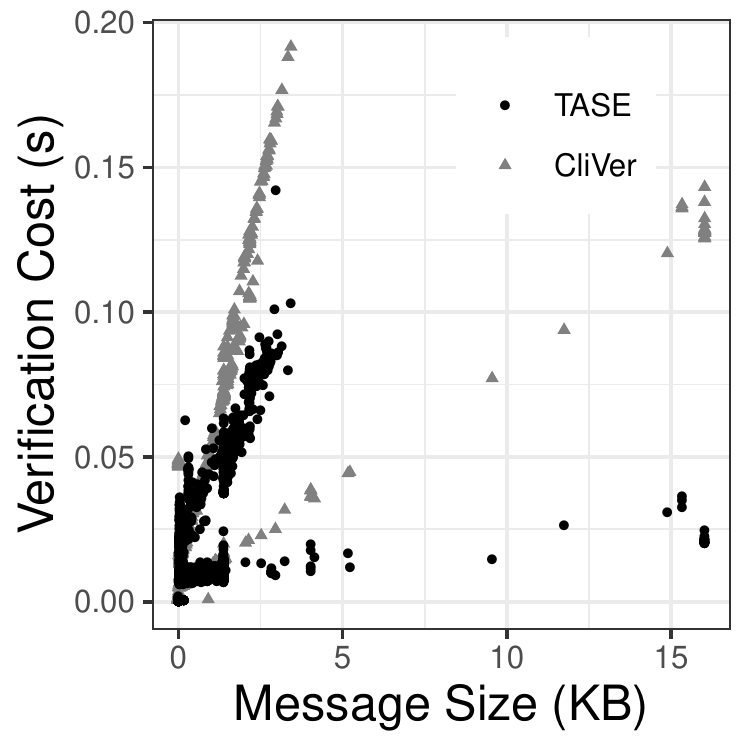}
    \caption{\textit{basic} configuration}
    \label{fig:costVsSize:basic}
  \end {subfigure}
  \begin{subfigure}[b]{.5\columnwidth}
    \includegraphics[width=\columnwidth]{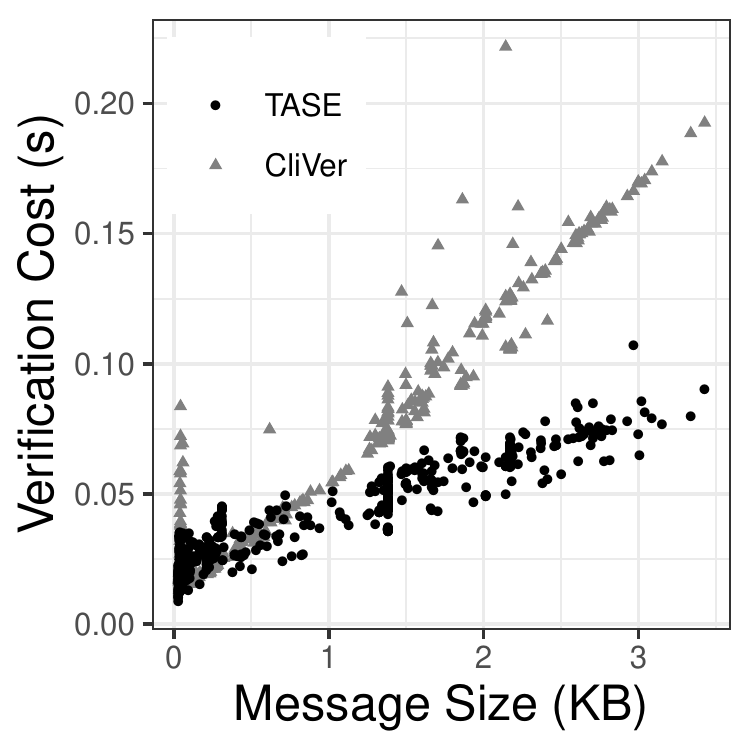}
    \caption{\textit{optimized} configuration}
     \label{fig:costVsSize:opt}
  \end{subfigure}
\caption{Verification cost vs.\ message size}
\label{fig:costVsSize}
\end{figure}

In \figref{fig:costVsSize} we report the \textit{cost} for verifying
each message in these connections as a function of the message's size.
The datapoints in \figref{fig:costVsSize} represent all 21 TLS
sessions in the \gmail dataset but, in the case of \cliver, omit
points for the ClientHello message and selected handshake messages of
each TLS connection.  These messages were omitted because \cliver's
excessive verification costs for them skewed the y-axis range
considerably, rendering the other trends more difficult to distinguish
visually.  (All messages are included in the \sysName datapoints,
however.)  \figref{fig:costVsSize:basic} represents the costs in the
\textit{basic} configuration, and \figref{fig:costVsSize:opt} shows
the costs in the \textit{optimized} configuration.  As can be seen in
\figref{fig:costVsSize:opt}, the costs for most messages scaled
linearly in message size for both \sysName and \cliver, but the slope
of this growth was flatter with \sysName, resulting in lower costs
(and so less lag) for verification.  In \figref{fig:costVsSize:basic},
the datapoints for \sysName fell along two lines corresponding to the
client-to-server and server-to-client messages (the latter are mostly
omitted from \figref{fig:costVsSize:opt}), and similarly for the
datapoints for \cliver.

\section{Limitations}
\label{sec:limitations}

In this section we discuss several limitations of \sysName.

\subsection{Equivalence of Interpretation and Native Execution}
\label{sec:limitations:equivalence}

Inevitable differences between the behaviors of native execution and
interpretation of the same project imply that \sysName results are not
necessarily semantically equivalent to those obtained using symbolic
execution based on interpretation only.  For example, \klee detects
out-of-bounds accesses to concrete buffers, while native execution
(without additional instrumentation) does not.  It would seem that
differences in the results of applying \klee and \sysName to a project
could arise, however, only due to the project's processing on
\textit{concrete} values, since processing on symbolic values would
trigger interpretation in \sysName, as well.
 
In the context of the client verification application discussed in
\secref{sec:eval:cliver}, this means that those behaviors permitted by
\cliver, which uses \klee to interpret the client program in full, are
not identical to those permitted by \sysName.  However, a behavior
permitted by \sysName but not by \cliver, if caused by an input
validation error of the client program, would presumably need to be an
artifact of server-to-client messages, which are concrete to the
verifier.  Since the verifier is deployed to defend the server and is
trusted to cooperate with it, malicious server-to-client messages are
outside the scope of those techniques.

\subsection{Interpreting x86 Instructions}
\label{sec:limitations:instructions}

Although the use of native state as the primary representation for
program execution in \sysName introduces opportunities for speculative
native execution on concrete data, this design choice also introduces
some difficulties.

Because \klee requires \llvm IR to perform symbolic execution, we
needed to produce \llvm IR models for the effects of each x86
instruction to be interpreted by \klee on the program's state.  In
addition to providing a burdensome engineering challenge, we found (as
did \sTwoE's authors~\cite{Chipounov:2011:S2E}) that modeling a given
x86 instruction's impact on program state using the RISC-like \llvm IR
required several \llvm IR instructions to fully capture all side
effects, including the changes to the FLAGS register. 

As a result, a machine-independent interpretation of a source program
in vanilla \klee could require fewer \llvm IR instructions to model
the program's execution than in \sysName.  We feel that our use cases
contain a sufficiently large usage of concrete data to justify the
optimizations for native execution in \sysName, but a tradeoff
nevertheless exists between the additional instructions needed for
interpretation in \sysName and the speed gained in native execution.

\subsection{Instrumentation}
In order to ensure that reads or writes to memory addresses containing
symbolic values are accurately recorded, \sysName uses a custom
\llvm backend to emit and instrument code.  Although \llvm provides 
many utilities for writing compiler passes to analyze or modify machine 
code as it is emitted, significant engineering challenges must still
be overcome to ensure that all code emitted for \sysName is properly
instrumented.  Specifically, the large number and variety of x86
instructions available combined with their side effects and implicit
operands make it difficult to write a catch-all compiler pass that
determines how an instruction touches memory.  Furthermore, determining
exactly where in the \llvm backend to inject instrumentation can be
nontrivial, given that \llvm applies a large number of
stages of optimization, some of which may
modify code emitted earlier during compilation.

To simplify the instrumentation process, \sysName's \llvm backend
restricts the pool of x86 instructions available to the compiler
during instruction selection.  Our anecdotal evidence suggests that
the slowdown imposed by choosing from a more limited set of
instructions is negligible compared to the overhead of setting up and
committing transactions and periodically interpreting when needed, but
we may expand the set of allowed instructions in the future.

\subsection{Controlled Forking}

\sysName was designed to use native forking to explore different
execution paths, each in a different process, in order to avoid the
overhead of software-based copy-on-write mechanisms as used in \klee
and \sTwoE~\cite{Cadar:2008:Klee, Chipounov:2011:S2E}.  Although
forking allows \sysName to explore distinct paths in parallel,
exploring distinct execution paths within distinct address spaces
complicates the process of applying search heuristics across these
many processes, sharing cached SMT query results across paths, etc.
Furthermore, even if it were desirable to move all or some aspects of
path exploration into a single address space, the TSX transactions
utilized for our speculative execution scheme would likely abort more
often due to their original intended use---detecting conflicting
concurrent accesses---thereby impinging on performance.

As discussed in \secref{sec:design:stateManagement}, our present
implementation leverages a central manager process to guide path
exploration, which it does simply by prioritizing which worker
processes it allows to proceed (and temporarily suspending others).
Some applications might require more sophisticated mechanisms for
state management in which this simple prioritization is insufficient.
For example, hybrid symbolic execution as introduced in
\mayhem~\cite{Cha:2012:Mayhem}, in which symbolic states can be
archived to relieve memory pressure and restored later for further
exploration, might be needed for analyzing some types of applications
efficiently.

\section{Conclusion}
\label{sec:conclusion}

In this paper, we presented the design, implementation, and evaluation
of \sysName.  To our knowledge, \sysName is the first symbolic
execution engine that leverages specialized hardware capabilities to
accelerate native execution to optimize workloads in which operations
on concrete data are a major bottleneck.  The two technical
innovations in \sysName to make this possible are (i) batching tests
to detect native accesses to symbolic data into a few instructions,
and (ii) undoing the potentially erroneous effects of having accessed
symbolic data natively by leveraging hardware transactions.

We illustrated an application of \sysName for verifying whether the
messaging behavior of a client as seen by the server is consistent
with the software the client is believed to be executing.  We showed
that the use of \sysName in this application dramatically reduced the
lag associated with verifying \openssl TLS 1.2 traffic, e.g., as
driven by Gmail.  This reduction bolsters the prospects of deploying
this verification on the critical path of delivering client messages
to the server, as an inline defense against client exploits without
foreknowledge of server vulnerabilities.

\section*{Acknowledgments}

This research was supported in part by grant N00014-17-1-2369
from the U.S.\ Office of Naval Research.

\bibliographystyle{IEEEtranS}
\bibliography{main}

\end{document}